\documentclass[showpacs,preprintnumbers,
amsmath,amssymb,aps,prd]{revtex4}
\usepackage{amsfonts}
\usepackage{amsfonts,graphics,epsfig,subfigure}

\begin{document}

\title{Coexistence curves and molecule number densities of AdS black holes in the reduced parameter space}
\author{Jie-Xiong Mo \footnote{mojiexiong@gmail.com}, Gu-Qiang Li \footnote{zsgqli@hotmail.com}}

 \affiliation{Institute of Theoretical Physics, Lingnan Normal University, Zhanjiang, 524048, Guangdong, China}

\begin{abstract}
   In this paper, we investigate the coexistence curves and molecule number densities of $f(R)$ AdS black holes and Gauss-Bonnet AdS black holes. Specifically, we work with the reduced parameter space and derive the analytic expressions of the universal coexistence curves that are independent of theory parameters. Moreover, we obtain the explicit expressions of the physical quantity describing the difference of the number densities of black hole molecules between the small and large black hole. It is found that both the coexistence curve and the difference of the molecule number densities of $f(R)$ AdS black holes coincide with those of RN-AdS black holes. It may be attributed to the same equation of state they share in the reduced parameter space. The difference of the molecule number densities between the small and large Gauss-Bonnet AdS black hole exhibits different behavior. This may be attributed to the fact that the charge of RN-AdS black hole is non-trivial. Our research will not only deepen the understanding of both the physics along the coexistence curve and the underlying microscopic freedom of AdS black holes, but also highlight the importance of the law of corresponding states.
\end{abstract}

\keywords{coexistence curve\;  black hole\; law of corresponding states}
  \maketitle

\section{Introduction}
    Ever since the pioneer work of Hawking and Page\cite{Hawking2}, the fascinating phase transition phenomena of AdS black holes have attracted the attention of researchers. Especially, it was found that charged AdS black holes behave much like the liquid-gas system. Chamblin et al.~\cite{Chamblin1,Chamblin2} first observed that the critical behavior of Reissner-Nordstr\"{o}m-AdS (RN-AdS) black holes is analogous to the Van der Waals (VdW) liquid-gas phase transition. Considering the variation of the cosmological constant in the first law~\cite{Caldarelli}-\cite{Lu}, thermodynamics of black holes in the extended space have been the focus recently. Treating the cosmological constant as thermodynamic pressure and its conjugate quantity as thermodynamic volume, the analogy of charged AdS black holes with the liquid-gas system has been further enhanced via $P-V$ criticality research~\cite{Kubiznak}-\cite{Shaowen3} initiated by Kubiznak and Mann~\cite{Kubiznak}. It was shown that for any nontrivial value of the charge, there exists a critical temperature below which the phase transition occur. There also exists a coexistence line terminating at the critical point. Critical exponents were calculated and found to coincide with the VdW fluid. Besides the VdW type first-order phase transition, much more interesting phenomena have been discovered and much richer physics has been revealed in the subsequent research. Reentrant phase transitions reminiscent of multicomponent liquids were discovered in Born-Infeld AdS black holes~\cite{Gunasekaran}, $d\geq6$ dimensional single spinning Kerr-AdS black holes~\cite{Altamirano1}, doubly-spinning Myers-Perry black holes~\cite{Altamirano3}, Gauss-Bonnet AdS black holes~\cite{Shaowen1} and Lovelock AdS black holes~\cite{Frassino}. Small/intermediate/large black hole phase transition reminiscent of solid/liquid/gas phase transition was reported in multiply rotating Kerr-AdS black holes~\cite{Altamirano2}. For nice reviews, see Ref.~\cite{Altamirano3,Dolan4}.

     Coexistence curves have been covered in literatures investigating $P-V$ criticality of black holes. However, in most of literatures, only the graphs of $P-T$ curves were presented. Recently, Zhao et al.~\cite{zhao1,zhao2} studied Clapeyron equations and parameter equations of coexistence curves. Utilizing a different approach, Wei et al.~\cite{Shaowen2} investigated the Clapeyron equations and fitting formula of the coexistence curve of RN-AdS black holes. It was found that the coexistence curve is charge independent in the reduced parameter space for any dimension of spacetime.

     Recently, Wei et al. introduce the number density of black hole molecules to measure the microscopic degrees of freedom of black holes~\cite{Shaowen3}. It was shown that the number density suffers a sudden change when black holes cross the coexistence curve while it is continuous at the critical point. Inspired by their findings, we would like to further probe in this paper whether their results can be generalized to black holes in modified gravity. For this purpose, we would like to investigate the coexistence curves and molecule number densities of $f(R)$ AdS black holes and Gauss-Bonnet AdS black holes.

    The organization of this paper is as follows. Coexistence curve and molecule number density of $f(R)$ AdS black holes will be studied in Sec. \ref{Sec2} while coexistence curve and molecule number density of Gauss-Bonnet AdS black holes will be investigated in Sec. \ref{Sec3}. A brief conclusion will be drawn in Sec. \ref {Sec4} .

\section{Coexistence curve and molecule number density of $f(R)$ AdS black holes}
\label{Sec2}

In this section, we study the case of $f(R)$ AdS black holes.

To start with, $P-V$ criticality of $f(R)$ AdS black holes will be briefly reviewed. Moon et al derived charged AdS black hole solution in the $R+f(R)$ gravity with Ricci scalar curvature $R=R_0$. The metric reads~\cite{Moon98}
\begin{equation}
ds^2=-N(r)dt^2+\frac{dr^2}{N(r)}+r^2(d\theta^2+sin^2\theta d\phi^2),\label{1}
\end{equation}%
where
\begin{eqnarray}
N(r)&=&1-\frac{2m}{r}+\frac{q^2}{br^2}-\frac{R_0}{12}r^2,\label{2}
\\
b&=&1+f'(R_0).\label{3}
\end{eqnarray}%
Note that the above solution is asymptotically AdS when the Ricci scalar curvature is treated as $R_0=-\frac{12}{l^2}=-4\Lambda$.

The parameters $m$ and $q$ are related to the black hole ADM mass $M$ and the electric charge $Q$ as follows
\begin{equation}
M=mb,\;\;\; Q=\frac{q}{\sqrt{b}}.\label{4}
\end{equation}%

Chen et al.~\cite{Chen} identified thermodynamic pressure as $P=-\frac{bR_0}{32\pi}$ and derived the equation of state and Gibbs free energy as follows
\begin{eqnarray}
P&=&\frac{bT}{v}-\frac{b}{2\pi v^2}+\frac{2q^2}{\pi v^4},\label{5}
\\
G&=&\frac{1}{4}\left(br_+-\frac{8\pi P r_+^3}{3}+\frac{3q^2}{r_+}\right). \label{6}
\end{eqnarray}%
They also obtained the critical point as~\cite{Chen}
\begin{equation}
T_c=\frac{\sqrt{6b}}{18\pi q},\;v_c=\frac{2\sqrt{6}\,q}{\sqrt{b}},\;P_c=\frac{b^2}{96\pi q^2}.\label{7}
\end{equation}%
They plotted $P-T$ curve and argued that "the coexistence line depends on the form of $f(R)$ and its slope increases with the derivative term $f'(R_0)$"~\cite{Chen}.

To calculate the critical exponents, they introduced the definitions as follows
\begin{equation}
p=\frac{P}{P_c},\;\nu=\frac{v}{v_c},\;\tau=\frac{T}{T_c},\label{8}
\end{equation}%
and rewrote the equation of state into~\cite{Chen}
\begin{equation}
p=\frac{8\tau}{3\nu}-\frac{2}{\nu^2}+\frac{1}{3\nu^4},\label{9}
\end{equation}%

Now we begin to investigate coexistence curve in the reduced parameter space.

Utilizing the definition $v=2r_+$, Eq. (\ref{6}) can be reexpressed as
\begin{equation}
G=\frac{1}{4}\left(\frac{bv}{2}-\frac{\pi P v^3}{3}+\frac{6q^2}{v}\right). \label{10}
\end{equation}%
Substituting Eq. (\ref{7}) into Eq.(\ref{10}), we obtain
\begin{equation}
G_c=\frac{\sqrt{6b}}{3}q, \label{11}
\end{equation}%
where $G_c$ denotes the Gibbs free energy at the critical point.

To study the behavior of Gibbs free energy in the reduced parameter space, we introduce the following definition
\begin{equation}
\tilde{G}=\frac{G}{G_c}. \label{12}
\end{equation}%
Utilizing Eqs.(\ref{7}), (\ref{8}), (\ref{10}), (\ref{11}) and (\ref{12}), one can derive
\begin{equation}
\tilde{G}=\frac{3+6\nu^2-p\nu^4}{8\nu}. \label{13}
\end{equation}%
From the above equation, one can see clearly that $\tilde{G}$ is independent of both the parameter $q$ and the parameter $b$ associated with the derivative term $f'(R_0)$.

Eq. (\ref{9}) can be reexpressed as
\begin{equation}
\tau=\frac{-1+6\nu^2+3p\nu^4}{8\nu^3}. \label{14}
\end{equation}%

Suppose there exist two phases at an arbitrary point of the coexistence curve. The physical quantities in the reduced parameter space are denoted as $\nu_1,\;p,\;\tau,\;\tilde{G}_1$ and $\nu_2,\;p,\;\tau,\;\tilde{G}_2$ respectively. According to Eqs. (\ref{13}) and (\ref{14}), one can easily write down the following equations
\begin{eqnarray}
\tilde{G}_1&=&\frac{3+6\nu_1^2-p\nu_1^4}{8\nu_1},\label{15}
\\
\tilde{G}_2&=&\frac{3+6\nu_2^2-p\nu_2^4}{8\nu_2},\label{16}
\\
\tau&=&\frac{-1+6\nu_1^2+3p\nu_1^4}{8\nu_1^3},\label{17}
\\
\tau&=&\frac{-1+6\nu_2^2+3p\nu_2^4}{8\nu_2^3}.\label{18}
\end{eqnarray}%

The black holes undergo first order phase transition along the coexistence curve. So the two phases mentioned above have the same Gibbs free energy, implying that $\tilde{G}_1$ =$\tilde{G}_2$. And Eqs. (\ref{15})-(\ref{18}) can be reorganized as
\begin{eqnarray}
\frac{3+6\nu_1^2-p\nu_1^4}{8\nu_1}&=&\frac{3+6\nu_2^2-p\nu_2^4}{8\nu_2},\label{19}
\\
\frac{-1+6\nu_1^2+3p\nu_1^4}{8\nu_1^3}&=&\frac{-1+6\nu_2^2+3p\nu_2^4}{8\nu_2^3},\label{20}
\\
2\tau&=&\frac{-1+6\nu_1^2+3p\nu_1^4}{8\nu_1^3}+\frac{-1+6\nu_2^2+3p\nu_2^4}{8\nu_2^3}.\label{21}
\end{eqnarray}%

\begin{figure*}
\centerline{\subfigure[]{\label{1a}
\includegraphics[width=8cm,height=6cm]{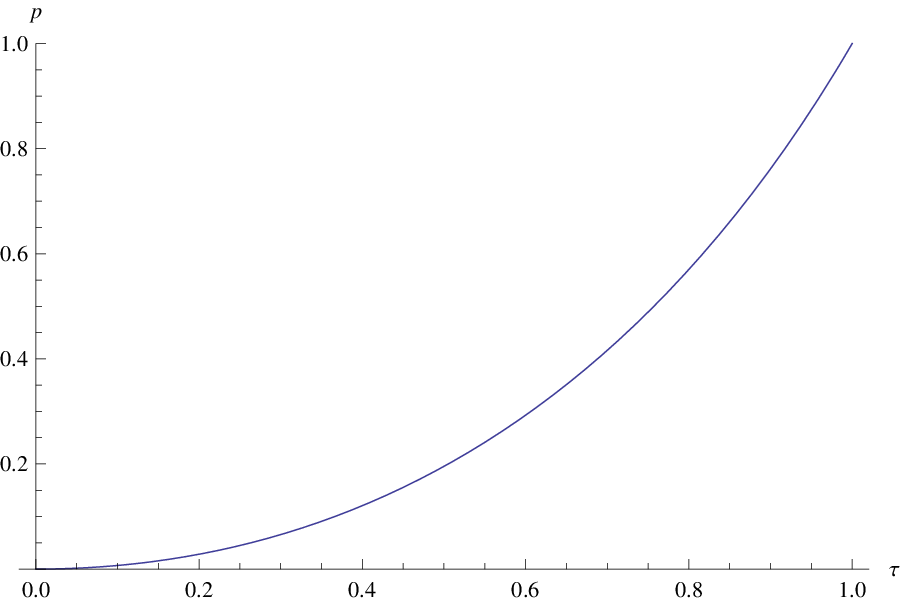}}
\subfigure[]{\label{1b}
\includegraphics[width=8cm,height=6cm]{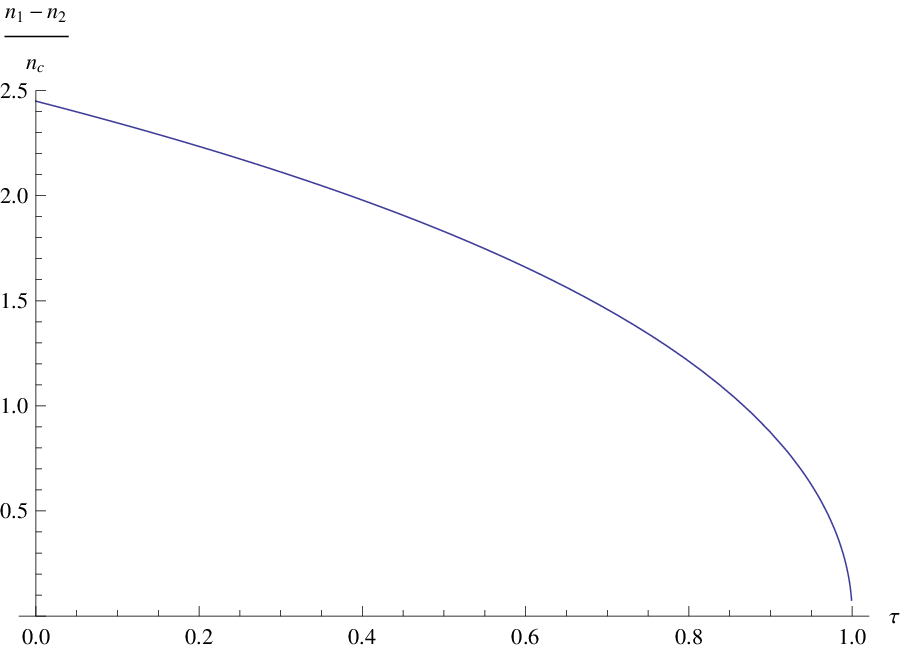}}}
 \caption{(a) coexistence curve of $f(R)$ AdS black holes in the reduced parameter space (b) the number density difference of the small and large $f(R)$ AdS black hole when $\tau \neq 1$} \label{fg1}
\end{figure*}

Introducing the definition that $\nu_1+\nu_2=x,\;\nu_1\nu_2=y$, the above three equations can be simplified into the following form when $\nu_1 \neq \nu_2$
\begin{eqnarray}
3-6y+py(x^2-y)&=&0,\label{22}
\\
x^2-y-6y^2+3py^3&=&0,\label{23}
\\
16\tau y^3-3py^3x-6y^2x+x(x^2-3y)&=&0.\label{24}
\end{eqnarray}%
Solving these equations, one can obtain
\begin{equation}
p=\frac{2\times 2^{1/3}(-\tau+\sqrt{-2+\tau^2})^{2/3}}{[2^{1/3}+(-\tau+\sqrt{-2+\tau^2})^{2/3}]^2}, \label{25}
\end{equation}%
from which one can safely draw the conclusion that the equation of the coexistence curve in the reduced parameter space is not only independent of the parameter $q$, but also independent of the parameter $b$ associated with the derivative term $f'(R_0)$. This can also be witnessed in Fig. \ref{1a}.

%%%%%%%%%%%%%%%%%%%%%%%%%%%%%%%%%%%%%%%%%%%%%%%%%%%%%%%%%%%%%%%%%%%%%%%%%%%%%
%%%%%%%%%%%%%%%%%%%%%%%%%%%%%%%%%%%%%%%%%%%%%%%%%%%%%%%%%%%%%%%%%%%%%%%%%%%%%%%%

Utilizing the definition of number density of black hole molecules $n=\frac{1}{v}$ introduced by Wei et al.~\cite{Shaowen3}, one can obtain
\begin{equation}
\frac{n_1-n_2}{n_c}=\frac{\frac{1}{v_1}-\frac{1}{v_2}}{\frac{1}{v_c}}=\frac{\nu_2-\nu_1}{\nu_1\nu_2}=\frac{\sqrt{x^2-4y}}{y}. \label{26}
\end{equation}%
The quantity $\frac{n_1-n_2}{n_c}$ describes the difference of the number densities of black hole molecules between the small and large black hole. Here we have made the assumption that $\nu_2>\nu_1$.
Substituting the solutions of Eqs. (\ref{22})-(\ref{24}) into Eq. (\ref{26}), we get
\begin{equation}
\frac{n_1-n_2}{n_c}=\sqrt{6-\frac{6\times 2^{2/3}\tau(-\tau+\sqrt{-2+\tau^2})^{1/3}}{2^{1/3}+(-\tau+\sqrt{-2+\tau^2})^{2/3}}}. \label{27}
\end{equation}%
Note that Eqs. (\ref{22})-(\ref{24}) are derived under the condition that $\nu_1 \neq \nu_2$. So Eq. (\ref{27}) holds for $\nu_1 \neq \nu_2$. At the critical point, $\nu_2=\nu_1=\nu_c$. So $\frac{n_1-n_2}{n_c}$ equals to zero when $\tau=1$.

Eq. (\ref{27}) is also independent of the parameter $b$ associated with the derivative term $f'(R_0)$. Eq. (\ref{27}) is shown intuitively in Fig. \ref{1b}, from which one can observe that the difference of the number densities between the small and large black hole decreases with the reduced temperature and approaches zero at the critical point. Comparing Fig.\ref{1a} and \ref{1b} with those of RN-AdS black holes in Refs. \cite{Shaowen2,Shaowen3}, one may find that they are exactly the same. It can be attributed to the same equation of state they share in the reduced parameter space.

\section{Coexistence curve and molecule number density of Gauss-Bonnet AdS black holes}
\label{Sec3}
In this section, we focus on Gauss-Bonnet AdS black holes.

To start with, we would briefly review the
$P-V$ criticality of Gauss-Bonnet AdS black holes. The action and the solution of $d$-dimensional Einstein-Maxwell theory with a Gauss-Bonnet term and a cosmological constant have been reviewed in Ref.~\cite{Cai98} as
\begin{eqnarray}
S&=&\frac{1}{16\pi}\int d^dx\sqrt{-g}\left[R-2\Lambda+\alpha_{GB}(R_{\mu\nu\gamma\delta}R^{\mu\nu\gamma\delta}-4R_{\mu\nu}R^{\mu\nu}+R^2)-4\pi F_{\mu\nu}F
^{\mu\nu}\right],\label{28}
\\
ds^2&=&-f(r)dt^2+f(r)^{-1}dr^2+r^2h_{ij}dx^idx^j
,\label{29}
\end{eqnarray}%
where
\begin{equation}
f(r)=k+\frac{r^2}{2\tilde{\alpha}}\left(1-\sqrt{1+\frac{64\pi \tilde{\alpha}M}{(d-2)\sum_kr^{d-1}}-\frac{2\tilde{\alpha}Q^2}{(d-2)(d-3)r^{2d-4}}-\frac{64\pi \tilde{\alpha}P}{(d-1)(d-2)}}\,\right).\label{30}
\end{equation}%
$\sum_k$ is the volume of the $(d-2)$ dimensional Einstein space whose line element reads $h_{ij}dx^idx^j$. $\alpha_{GB}$ is the Gauss-Bonnet coefficient and $\tilde{\alpha}=(d-3)(d-4)\alpha_{GB}$.

Solving the equation $f(r)=0$ for the largest root, one can obtain the horizon radius $r_h$. And the mass can be expressed as
\begin{equation}
M=\frac{(d-2)\sum_kr_h^{d-3}}{16\pi}\left(k+\frac{k^2\tilde{\alpha}}{r_h^2}+\frac{16\pi P r_h^2}{(d-1)(d-2)}\right)+\frac{\sum_kQ^2}{8\pi (d-3)r_h^{d-3}}.\label{31}
\end{equation}%

The corresponding Hawking temperature and entropy were derived as~\cite{Cai98}
\begin{eqnarray}
T&=&\frac{f'(r_h)}{4\pi}=\frac{16\pi Pr_h^4/(d-2)+(d-3)kr_h^2+(d-5)k^2\tilde{\alpha}-\frac{2Q^2}{(d-2)r_h^{2d-8}}}{4\pi r_h(r_h^2+2k\tilde{\alpha})},\label{32}
\\
S&=&\frac{\sum_kr_h^{d-2}}{4}\left(1+\frac{2(d-2)\tilde{\alpha}k}{(d-4)r_h^2}\right). \label{33}
\end{eqnarray}%

Identifying the thermodynamic pressure as $P=-\frac{\Lambda}{8\pi}$, the first law of black hole thermodynamics and the corresponding Smarr relation in the extended phase space take new forms as~\cite{Cai98}
\begin{eqnarray}
dH&=&TdS+\Phi dQ+VdP+\mathcal{A} d\tilde{\alpha},\label{34}
\\
(d-3)H&=&(d-2)TS-2PV+2\mathcal{A}\tilde{\alpha}+(d-3)Q\Phi, \label{35}
\end{eqnarray}%
where $\mathcal{A}$ is the quantity conjugate to $\tilde{\alpha}$.

When $Q=0$, the equation of state reads
\begin{equation}
P=\frac{d-2}{4r_h}(1+\frac{2k\tilde{\alpha}}{r_h^2})T-\frac{(d-2)(d-3)k}{16\pi r_h^2}-\frac{(d-2)(d-5)k^2\tilde{\alpha}}{16\pi r_h^4}.\label{36}
\end{equation}%

It was disclosed in Ref.~\cite{Cai98} that all the three physical quantities at the critical point can be analytically solved when $Q=0,k=1,d=5$. So in this paper, we would mainly concentrate on the five-dimensional uncharged Gauss-Bonnet AdS black holes. The corresponding critical point was obtained as~\cite{Cai98}
\begin{equation}
r_{hc}=\sqrt{6\tilde{\alpha}},\;T_c=\frac{1}{\pi \sqrt{24\tilde{\alpha}}},\;P_c=\frac{1}{48\pi \tilde{\alpha}}.\label{37}
\end{equation}%

Now we begin to investigate coexistence curve in the reduced parameter space.

The specific volume was identified as $v=\frac{4r_h}{d-2}$~\cite{Cai98}. When $d=5$, $v=\frac{4r_h}{3}$. Utilizing this relation and substituting $d=5, k=1$ into Eq. (\ref{36}), one can obtain
\begin{equation}
P=\frac{T}{v}+\frac{32\tilde{\alpha}T}{9v^3}-\frac{2}{3\pi v^2}.\label{38}
\end{equation}%

The Gibbs free energy is defined as
\begin{equation}
G=H-TS=M-TS.\label{39}
\end{equation}%
Note that the mass $M$ should be interpreted as the enthalpy in the extended phase space and we have utilized the identification $H=M$ in the above derivation.

Substituting Eqs. (\ref{31}), (\ref{33}) into Eq.(\ref{39}) and utilizing Eq.(\ref{38}), one can obtain the explicit expression of Gibbs free energy of five-dimensional uncharged Gauss-Bonnet AdS black holes as follow
\begin{equation}
G=\frac{3\sum_k}{1024\pi}(64\tilde{\alpha}+18v^2-9\pi T v^3-288\pi \tilde{\alpha}Tv).\label{40}
\end{equation}%

Utilizing the definitions in Eq. (\ref{8}), Eqs. (\ref{38}) and (\ref{40}) can be rewritten as
\begin{eqnarray}
\tau&=&\frac{\nu (3+p\nu^2)}{1+3\nu^2},\label{41}
\\
G&=&\frac{3\tilde{\alpha}\sum_k}{16\pi}[1+3\nu^2-\nu \tau (3+\nu^2)].\label{42}
\end{eqnarray}%

Suppose there exist two phases at an arbitrary point of the coexistence curve. The corresponding physical quantities are denoted as $\nu_1,\;p,\;\tau,\;G_1$ and $\nu_2,\;p,\;\tau,\;G_2$ respectively. According to Eqs. (\ref{41}) and (\ref{42}), one can easily write down the following equations
\begin{eqnarray}
G_1&=&\frac{3\tilde{\alpha}\sum_k}{16\pi}[1+3\nu_1^2-\nu_1 \tau (3+\nu_1^2)],\label{43}
\\
G_2&=&\frac{3\tilde{\alpha}\sum_k}{16\pi}[1+3\nu_2^2-\nu_2 \tau (3+\nu_2^2)],\label{44}
\\
\tau&=&\frac{\nu_1 (3+p\nu_1^2)}{1+3\nu_1^2},\label{45}
\\
\tau&=&\frac{\nu_2 (3+p\nu_2^2)}{1+3\nu_2^2}.\label{46}
\end{eqnarray}%

The two phases mentioned above share the same Gibbs free energy because the black holes undergo first order phase transition along the coexistence curve. And Eqs. (\ref{43})-(\ref{46}) can be reorganized as
\begin{eqnarray}
\frac{3\tilde{\alpha}\sum_k}{16\pi}[1+3\nu_1^2-\nu_1 \tau (3+\nu_1^2)]&=&\frac{3\tilde{\alpha}\sum_k}{16\pi}[1+3\nu_2^2-\nu_2 \tau (3+\nu_2^2)],\label{47}
\\
\frac{\nu_1 (3+p\nu_1^2)}{1+3\nu_1^2}&=&\frac{\nu_2 (3+p\nu_2^2)}{1+3\nu_2^2},\label{48}
\\
2\tau&=&\frac{\nu_1 (3+p\nu_1^2)}{1+3\nu_1^2}+\frac{\nu_2 (3+p\nu_2^2)}{1+3\nu_2^2}.\label{49}
\end{eqnarray}%

Introducing the definition that $\nu_1+\nu_2=x,\;\nu_1\nu_2=y$, the above three equations can be simplified into the following form when $\nu_1 \neq \nu_2$
\begin{eqnarray}
3x-3\tau-\tau (x^2-y)&=&0,\label{50}
\\
3+p(x^2-y)-9y+3py^2&=&0,\label{51}
\\
2\tau (1+3x^2+9y^2-6y)-3x-9xy-px(x^2-3y)-3pxy^2&=&0.\label{52}
\end{eqnarray}%

\begin{figure*}
\centerline{\subfigure[]{\label{2a}
\includegraphics[width=8cm,height=6cm]{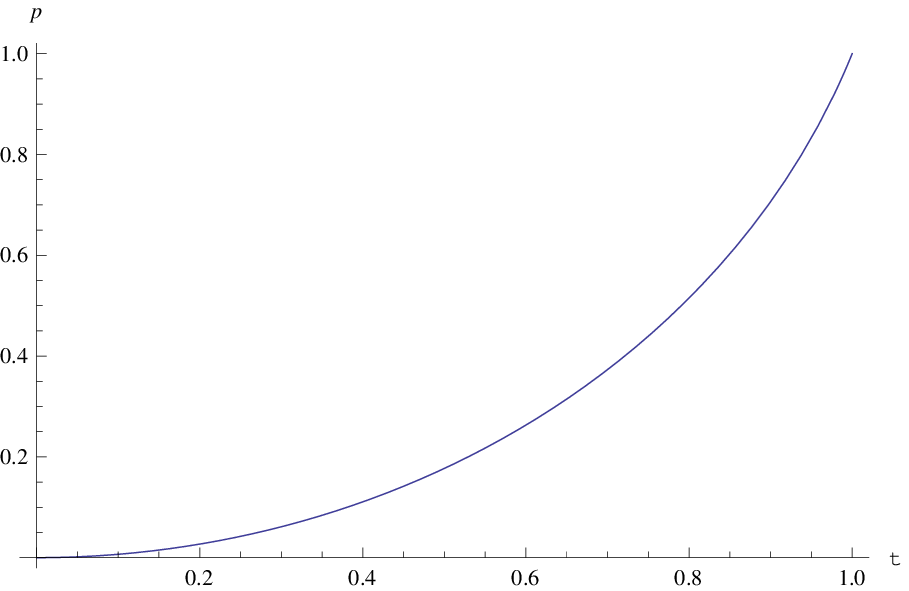}}
\subfigure[]{\label{2b}
\includegraphics[width=8cm,height=6cm]{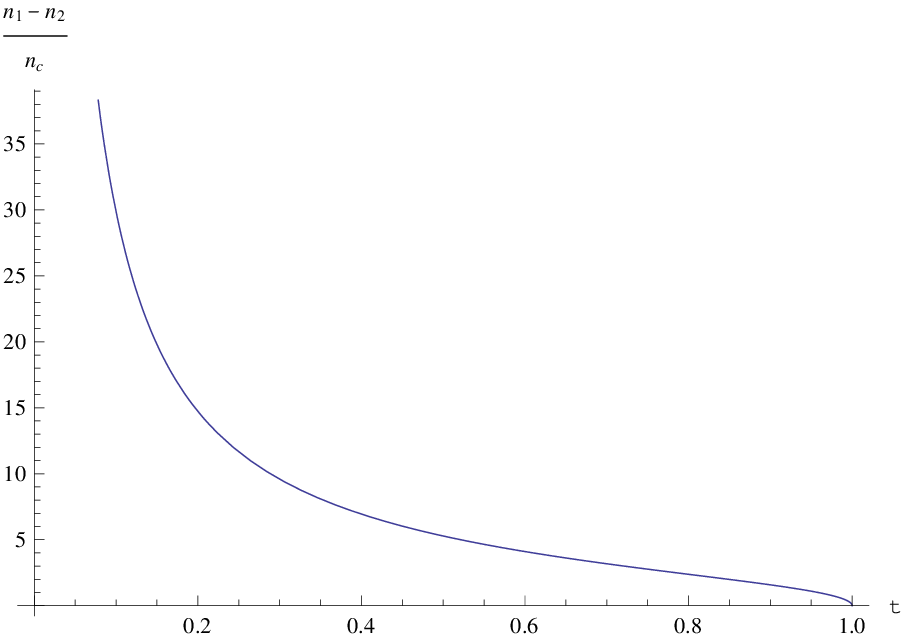}}}
 \caption{(a) coexistence curve of Gauss-Bonnet AdS black hole in the reduced parameter space (b) the number density difference of the small and large Gauss-Bonnet AdS black hole when $\tau \neq 1$} \label{fg2}
\end{figure*}

Solving these equations, one can obtain
\begin{equation}
p=\frac{1}{2}(3-\sqrt{9-8\tau^2}), \label{53}
\end{equation}%
from which one can safely draw the conclusion that the equation of the coexistence curve in the reduced parameter space is independent of the parameter $\tilde{\alpha}$. This can also be witnessed in Fig. \ref{2a}.

%%%%%%%%%%%%%%%%%%%%%%%%%%%%%%%%%%%%%%%%%%%%%%%%%%%%%%%%%%%%%%%%%%%%%%%%%%%%%
%%%%%%%%%%%%%%%%%%%%%%%%%%%%%%%%%%%%%%%%%%%%%%%%%%%%%%%%%%%%%%%%%%%%%%%%%%%%%%%%

Utilizing the definition in Eq. (\ref{26}), one can obtain
\begin{equation}
\frac{n_1-n_2}{n_c}=\sqrt{-6+\frac{3(3+\sqrt{9-8\tau^2})}{2\tau^2}}. \label{54}
\end{equation}%
where the quantity $\frac{n_1-n_2}{n_c}$ describes the difference of the number densities of black hole molecules between the small and large black hole. Here we have made the assumption that $\nu_2>\nu_1$.

Note that Eqs. (\ref{50})-(\ref{52}) are derived under the condition that $\nu_1 \neq \nu_2$. So Eq. (\ref{54}) holds for $\nu_1 \neq \nu_2$. At the critical point, $\nu_2=\nu_1=\nu_c$. So $\frac{n_1-n_2}{n_c}$ equals to zero when $\tau=1$.

Eq. (\ref{54}) is also independent of the parameter $\tilde{\alpha}$. Eq. (\ref{54}) is shown intuitively in Fig. \ref{2b}, from which one can observe that the difference of the number densities between the small and large black hole decreases with the reduced temperature and approaches zero at the critical point. However, when $\tau$ approaches zero, the difference of the number densities between the small and large Gauss-Bonnet AdS black hole approaches infinity. This behavior is quite different from that of RN-AdS black hole. 

\section{Conclusions and discussions}
\label{Sec4}
 The coexistence curves and molecule number densities of $f(R)$ AdS black holes and Gauss-Bonnet AdS black holes are investigated in the reduced parameter space. We derive the analytic expressions of the universal coexistence curves that are independent of theory parameters. We further obtain the explicit expressions of the physical quantity describing the difference of the number densities of black hole molecules between the small and large black hole. It is found that both the coexistence curve and the difference of the molecule number densities of $f(R)$ AdS black holes coincide with those of RN-AdS black holes. It may be attributed to the same equation of state they share in the reduced parameter space. 
 
 However, the difference of the molecule number densities between the small and large Gauss-Bonnet AdS black hole exhibits different behavior. This difference can be attributed to the fact that the charge of RN-AdS black hole is non-trivial. And the horizon radius is limited from below by the inner horizon radius of the extremal black hole. So the difference of the molecule number densities between the small and large RN-AdS black hole is still finite when $\tau$ approaches zero. On the other hand, there is no lower limit on the horizon radius for the uncharged Gauss-Bonnet AdS black hole. So the difference of the molecule number densities between the small and large Gauss-Bonnet AdS black hole blows up to infinity when $\tau$ approaches zero. It is believed that the behavior for the charged Gauss-Bonnet AdS black hole case would be much more similar as that of the RN-AdS black hole. However, the expressions of critical quantities for the charged Gauss-Bonnet AdS black hole are quite lengthy and difficult to be simplified. Further investigation on the reduced parameter space of the charged Gauss-Bonnet AdS black hole is called for.
 
 Our research will not only deepen the understanding of both the physics along the coexistence curve and the underlying microscopic freedom of AdS black holes, but also highlight the importance of the law of corresponding states. In classical thermodynamics, "this law is valid under more general assumptions than the original derivation of the Van der Waals equation" \cite{Kubiznak}.

 \section*{Acknowledgements}
The authors would like to express their sincere gratitude to the referee whose deep physical insight has help interpreted the phenomenon we found in our research. This research is supported by Department of Education of Guangdong Province of China(Grant No.2014KQNCX191). It is
also supported by \textquotedblleft Thousand Hundred
Ten\textquotedblright \,Project of Guangdong Province.

\end{document}